\documentclass[fleqn,10pt]{wlscirep}
\usepackage[utf8]{inputenc}
\usepackage[T1]{fontenc}
 
 \usepackage{amsthm}
 
\newcommand{\updt}[1]{{\color{black}#1}}
  
\usepackage{verbatim}
\usepackage{setspace}
 
 \usepackage[numbers]{natbib}
\newtheorem*{Proof*}{Proof}

\newtheorem{Theorem}{Theorem}
\newtheorem{Example}{Example}

\newtheorem{Proposition}{Proposition}
\newcommand{\calK}{{\cal K}}

\newcommand{\p}[1]{\left(#1\right)}

\newcommand {\pr} {\mathbb{P}}
\newcommand{\pp}[1]{\left[#1\right]}
\newcommand{\ppp}[1]{\left\{#1\right\}}
\newcommand{\abs}[1]{\left|#1\right|}

\newcommand{\LMD}{\lambda_0,\dots,\lambda_n}
\newcommand{\s}[1]{\mathsf{#1}}

\newcommand {\R}{\mathbb R}

\newcommand{\be}{\begin{equation}}
\newcommand{\ee}{\end{equation}}

\title{Variability  in mRNA Translation: A Random Matrix Theory Approach}

\author[1,+]{Michael Margaliot}
\author[1,+]{Wasim Huleihel} 
\author[2,*]{Tamir Tuller} 
 
\affil[1]{ Department  of Electrical Engineering-Systems, Faculty of Engineering,  Tel Aviv University, Tel Aviv, Israel, 69978. }

\affil[2]{Department  of Biomedical Engineering, Faculty of Engineering,  Tel Aviv University, Tel Aviv, Israel, 69978. }

\affil[*]{tamirtul@post.tau.ac.il}
\affil[+]{These authors contributed equally to this work}

\begin{abstract}
  The rate of  mRNA translation depends on the initiation, elongation, and termination rates of ribosomes along the mRNA.  These rates depend on many ``local'' factors like the abundance of free ribosomes and tRNA molecules in the vicinity of the mRNA molecule. All these factors are stochastic  and their experimental measurements are also noisy.
 \updt{ An important question is how protein production in the cell is affected by this considerable variability.}
  We develop a new theoretical framework for addressing this question by \updt{modeling the rates as identically and independently  distributed random variables and using tools from random matrix theory to analyze the steady-state production rate.  The analysis reveals a principle of universality:  the average protein production rate depends only on the of the set of possible values that  the  random variable may attain. }
This  explains how total protein production can be stabilized despite the overwhelming  stochasticticity  
  underlying cellular processes.
\end{abstract}

\begin{document}

\flushbottom
\maketitle
\thispagestyle{empty}


\section{Introduction} 

During  translation  complex molecular machines called
 ribosomes  scan the mRNA codon by codon.	
The   ribosome  
 links amino-acids together in the order specified by the codons  to form a  polypeptide chain.   
 For each codon, the ribosome ``waits'' for
a   transfer RNA~(tRNA) molecule that matches and carries the correct amino-acid for  incorporating
it  into the  growing polypeptide chain. 
When the 
  ribosome reaches  a stop codon encoding a termination signal, it   detaches from the mRNA and  the complete  amino-acid chain is released.

	Several ribosomes may read the same mRNA molecule simultaneously, 
	as this form of ``pipelining'' increases the protein production  rate.
	The dynamics of   ribosome flow along the mRNA  strongly affects the   production rate and the correct folding of the protein. A ribosome  that is  stalled for a long time   may   lead to the formation
of a  ``traffic jam'' of ribosomes behind it, and consequently to depletion of the pool of free ribosomes.    
	Mutations affecting the protein translation rates may be  associated with various diseases~\cite{Kimchi-Sarfaty2011}, as well as  viral infection efficiency~\cite{Mioduser2017}.

As translation is a central metabolic process that consumes most of the energy in the cell~\cite{Lane2010,Mahalik2014,Buttgereit1995,Russell1995,Gorochowski2016}, 
cells operate  sophisticated regulation mechanisms to
avoid and resolve ribosome traffic jams~\cite{jusz2020,jusz2020dis,Millseaan2755,Tuller2010}. 
Another testimony of the importance of ribosome flow is the fact that
about half of the currently existing antibiotics target the bacterial ribosome by interfering with translation initiation,
elongation, termination and other regulatory mechanisms~\cite{ribo_cancer16,riboanti}.
For example,
Aminoglycosides inhibit bacterial protein synthesis by
  binding to the 30S ribosomal subunit,  
  stabilizing  a normal mismatch in codon–anticodon
pairing, and leading to mistranslations~\cite{anti_ribo}.  
Understanding 
  the mechanisms   of ribosome-targeting antibiotics and the molecular mechanisms of bacterial resistance
	is crucial for developing new  drugs that can effectively
	inhibit the synthesis of bacterial proteins~\cite{anti_ribo_rev}.

\begin{figure*}[t]
 \begin{center}
\includegraphics[width=10cm,height=8cm]{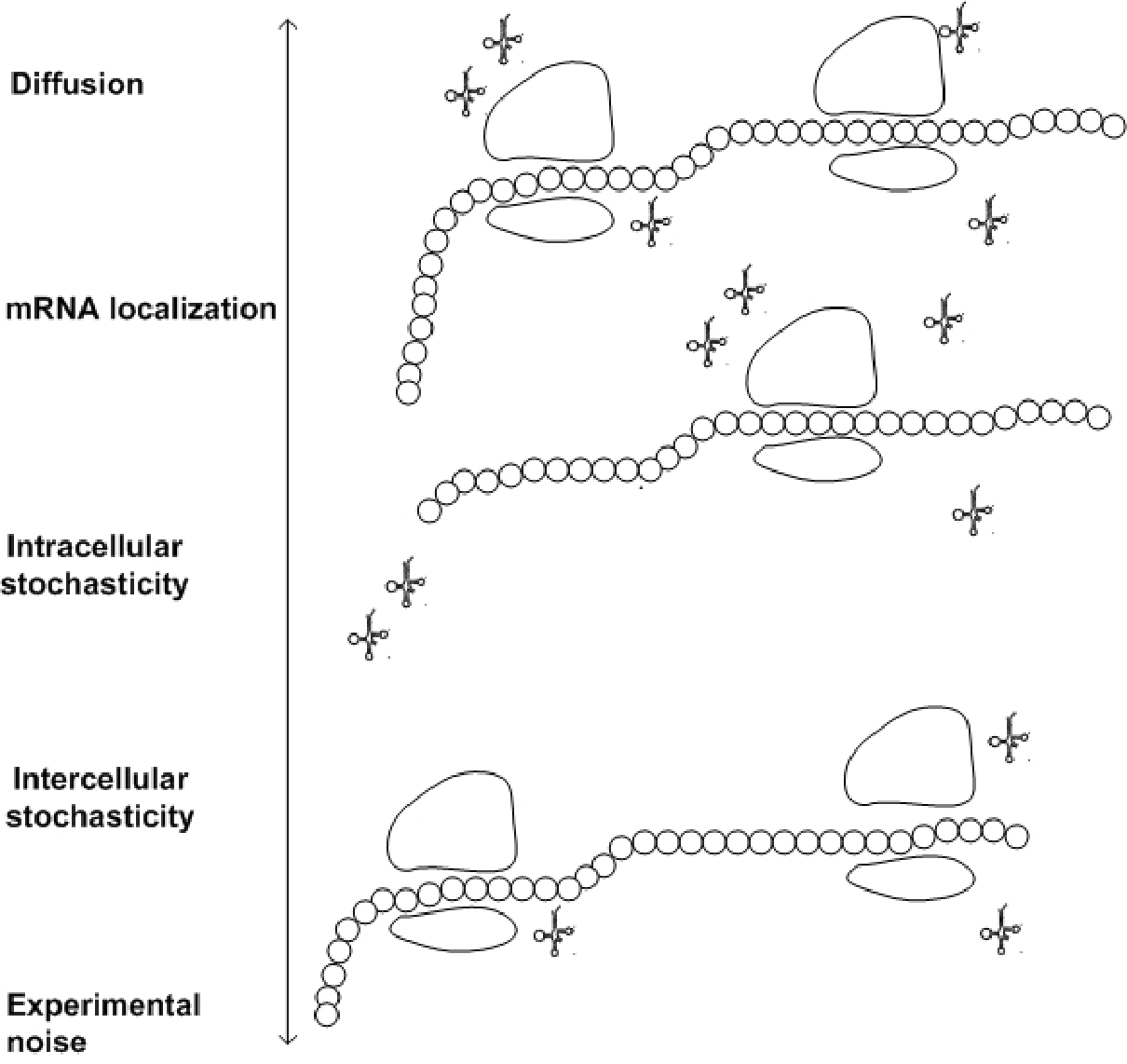}
\caption{Stochasticity and noise in mRNA translation and its measurements imply \updt{ that identical mRNAs chains may  have different transition rates. The double arrows   represent tRNA molecules.  }  }\label{fig:srfm}
\end{center}
\end{figure*}

Summarizing, an important  problem is to understand   the 
	  dynamics of   ribosome  flow along the mRNA, and how it 
	 affects the  protein   production rate. 
As in many  cellular processes, it is important to understand
how   proper functioning is maintained, and adjusted to  the signals that a cell
receives and to resource availability, 
in spite of the 
large stochasticity  in the cell~\cite{noise2003,noise_single}.
Translation and the measurements of this process are  affected by various types of stochasticity (see a review in~\cite{Sonneveld2020}), as illustrated in Figure~\ref{fig:srfm}. Specifically,
\begin{itemize}
    \item All the chemical reactions related to the process are of course stochastic, and
    so are 
    the concentrations of factors like cognate tRNA availability and the resulting translation rates (e.g. during cell cycle),
    structural accessibility of the $5'$-end to translation factors,   the spatial organization of mRNAs inside
    the cell \updt{and the existence of designated ``translation factories''}~\cite{Korkmazhan13424,LECUYER2007174,local_mrna2008,Sabi2019}.
    \item Different cells in a population are not identical for example in terms of the number of mRNA molecules and ribosomes in the cell and many  other aspects~\cite{Buettner2015}. 
    \item It was recently suggested that the ribosomes themselves are not identical~\cite{Genuth2019}. 
    \item The stochastic diffusion of translation substrates play a key role in determining translation rates~\cite{diff2019}. The fact that the mRNA molecules of the same gene diffuse (either actively or passively) to different regions in the cell   affects their translation properties~\cite{Martin2009}. 
    \item The experimental approaches for measuring translation introduce various types of noise~\cite{Gerashchenko2017,Diament2016}. Thus, the parameters of translation that are inferred from these data are also noisy.
     \item Processes such as mRNA methylation can affect all
     aspects of translation~\cite{Zaccara2019,Sonneveld2020}.
     \item There are couplings between the translation process and other stochastic 
     gene expression steps~\cite{Sonneveld2020,Bergman2020} such as
     transcription~\cite{McGary2013}, mRNA stability~\cite{Edri2014B,Presnyak2015}, and interaction with miRNA~\cite{Bazzini2012,Bergman2020A} and RNA binding proteins \cite{Sonneveld2020}.   
     
\end{itemize}

 \updt{A recent paper analyzes translation and concludes  that ``randomness, on average, plays a greater role than any non-random contributions to synthesis time''~\cite{PhysRevE.97.022409}.}

Here, we develop a   theoretical approach to analyze  
translation subject to spatial variation  by combining a  deterministic 
computational model, called the ribosome flow model~(RFM), with tools from  random matrix theory. We model the variation in the \updt{initiation, elongation, and exit} rates in several copies of the same mRNA by assuming that the rates in the RFM are \updt{independent and identically distributed}~(i.i.d.) random variables,   that is,  each random variable has the same probability distribution as the others and all are mutually independent .
\updt{
This assumption is of course
restrictive, and is needed to obtain our closed-form theoretical  results. 
  Yet, it seems to have some empirical justification. 
For example,  away from the ends of the coding sequence the translation rates tend to be independent~\cite{Tuller2015}. In addition, various  noise sources (such as   NGS noise) tend to be independent along the mRNA.
 }
\updt{Furthermore, in Section~\ref{sec:gen}  we describe several
generalizations   where the i.i.d. assumption  on the random variables can be  relaxed.} 

We believe that our approach can be used to tackle   various levels of stochaticity and uncertainty in translation and its measurements.
Our main results (Theorems~\ref{thm:main} and~\ref{thm:main_with_more} below) 
reveal a new principle  of universality: as the length of the mRNA molecule increases the overall  steady-state protein production rate 
converges, with probability one,  to a constant value that depends only on the minimal
possible value of the~random variables.
Roughly speaking, this suggests that much of the 
variability is ``filtered out'', 
and this may explain how the cell 
overcomes the variations  in the many stochastic factors mentioned above.  
 
The next section reviews  the~RFM and 
 some of  its  dynamical properties 
  that are relevant in our context. This is followed by our    theoretical results. Section~\ref{sec:gen} describes two generalizations. 
  The final section concludes and  describes several possible 
  directions for further research.

\section{ Ribosome Flow Model  (RFM) }\label{sec:rfm}

Mathematical models of the flow of ``biological particles'' like 
RNA polymerase, ribosomes, and molecular motors 
 are becoming increasingly  important,
 as powerful  experimental techniques provide rich data
on the dynamics  of such machines inside the cell~\cite{ribosome_profiling_rev,JCP:JCP24445,Mayer2016}, sometimes in real-time~\cite{Iwasaki1391}.
  Computational  models are particularly important in  fields like 
	synthetic biology and biotechnology, as they can provide 
		qualitative and quantitative  testifiable   predictions 
		on  the effects of  various manipulations  of the genetic machinery~\cite{Haar2012}.  They are also helpful for understanding the evolution of cells and their biophysics~\cite{Zur2016}. 

The standard computational 
 model for the flow of  biological particles 
 is the \emph{asymmetric simple exclusion process} (ASEP)~\cite{MacDonald1968,MacDonald1969, Spitzer1970246,TASEP_tutorial_2011,Shaw2003}. This  is a  fundamental model from nonequilibrium statistical mechanics
  describing particles that  hop randomly  from a site to a neighboring site
along an ordered (usually 1D) lattice. 
 Each site may be either free or occupied by a single particle, and hops may take place only to a free target site, representing the fact that the particles have volume and cannot overtake one another.
This simple exclusion principle 
generates an indirect coupling
 between the moving particles. The motion is assumed to be directionally asymmetric, i.e., there is some preferred direction of motion. In the \emph{totally asymmetric simple exclusion process} (TASEP) the motion is unidirectional.

 TASEP and its  variants have been used extensively 
to model and analyze   natural and artificial  processes
including ribosome  flow, vehicular and pedestrian traffic, molecular motor traffic,   the movement of ants along a trail, and more~\cite{TASEP_book,pinkoviezky2013transport,Zur2016}. However, due to the intricate
indirect interactions between the hopping particles,
analysis of TASEP is difficult, and closed-form  results exist only in some special 
cases~\cite{Derrida92,exactsol}.

\begin{figure*}[t]
 \begin{center}
\includegraphics[width=13cm,height=2.8cm]{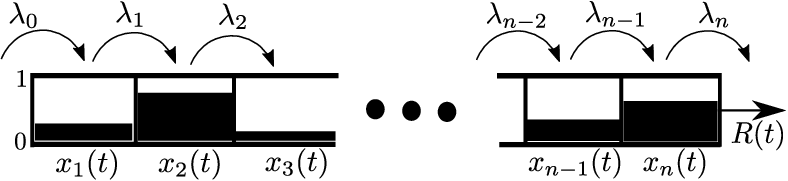}
\caption{Unidirectional flow along an~$n$ site    RFM. 
State variable~$x_i(t)\in[0,1]$ represents
 the normalized density at site~$i$ at time $t$. The parameter  $\lambda_i>0$ controls  the transition rate from  site~$i$ to site~$i+1$, with~$\lambda_0 $ [$\lambda_n $] controlling    the initiation [termination] rate.   $ {R} (t)$ is the output rate from the chain
at time~$t$. }\label{fig:rfm}
\end{center}
\end{figure*}

The  RFM~\cite{reuveni} is a  deterministic,
 nonlinear, continuous-time ODE  model  
 that can be derived via a dynamic
mean-field approximation of~TASEP~\cite{Zarai20170128}. 
It is amenable to rigorous  analysis using tools from systems and control theory. 
The RFM  includes~$n$   sites  ordered along a 1D chain.
 The normalized density (or occupancy level) of site~$i$ at time~$t$
    is described
		by a state variable~$x_i(t)$ that takes values in the interval~$[0,1]$,
			where~$x_i(t)=0$ [$x_i(t)=1$] represents  that site~$i$ is completely free [full] at time $t$.  
The transition between sites~$i$ and site~$i+1$
is regulated by a parameter~$\lambda_i>0$.
In particular,   $\lambda_0$ [$\lambda_n$] controls the initiation [termination] rate into [from] the chain.
 The rate at which particles exit the chain at time~$t$  is a scalar denoted by~${R}(t)$ (see Fig.~\ref{fig:rfm}).

When modeling the flow of biological machines  like  ribosomes 
the  chain models an
 mRNA  molecule coarse-grained into~$n$ sites. Each site is a      codon  or a group  of consecutive 
codons, and
  $ {R} (t)$
  is the rate at which  ribosomes detach from the mRNA, i.e. the protein production rate.
The values of the~$\lambda_i$s  encapsulate 
many  biophysical properties like
 the number of available free ribosomes,
the  nucleotide context surrounding initiation codons,
 the codon compositions in
 each site and the corresponding tRNA availability, 
and so on~\cite{reuveni,TullerGB2011,Dana2011}.
Note that these factors may vary in different locations inside the cell.

The dynamics of the RFM   is described  by $n$ nonlinear first-order ordinary differential equations:
\be\label{eq:rfm_all}
\dot{x}_i=\lambda_{i-1}x_{i-1}(1-x_i)-\lambda_i x_i(1-x_{i+1}),\quad i=1,\dots,n,
\ee
where we define~$x_0(t) := 1$ and $x_{n+1}(t) : = 0$.
Every~$x_i$ is dimensionless, and every rate~$\lambda_i$ has units of~$1/{\text{time}}$.
Eq.~\eqref{eq:rfm_all}    can be explained as follows. The flow of particles from site~$i$ to site~$i+1$ is~$\lambda_{i} x_{i}(t)
(1 - x_{i+1}(t) )$. This flow is proportional to~$x_i(t)$, i.e. it increases
with the occupancy level at site~$i$, and to $(1-x_{i+1}(t))$, i.e. it decreases as site~$i+1$ becomes fuller.  This is a ``soft''
 version of the simple  exclusion principle.
The maximal possible  flow  from site~$i$ to site~$i+1$  is the transition rate~$\lambda_i$.
 Eq.~\eqref{eq:rfm_all}   is thus a simple balance law:    the change in the  density~$x_i$  equals the  flow entering site~$i$ from site~$i-1$,     minus the flow exiting from site~$i$ to site~$i+1$.
 The output rate from  the last site at time~$t$ is~$ {R} (t):=\lambda_n x_n(t)$.

An important property of the RFM (inherited from~TASEP) is that it can be used to model and analyze the formation of ``traffic jams'' of particles along the chain. It was shown that traffic jams during translation are common phenomena even under  standard conditions~\cite{Diament2018}.
Indeed, suppose that there exists an index~$j$ such that~$\lambda_j$ is much smaller than all the other rates. Then~\eqref{eq:rfm_all} gives
\begin{align*} 
\dot{x}_j &= \lambda_{j-1}x_{j-1}(1-x_j)-\lambda_j x_j(1-x_{j+1})\\
          &\approx \lambda_{j-1}x_{j-1}(1-x_j),
 \end{align*}
this term is positive when~$x\in(0,1)^n$, 
so we can expect site~$j$ to fill up, i.e.~$x_j(t) \to 1$. Now using~\eqref{eq:rfm_all} again
gives
\begin{align*} 
\dot{x}_{j-1} &= \lambda_{j-2}x_{j-2}(1-x_{j-1})-\lambda_{j-1} x_{j-1} (1-x_{j})\\
          &\approx \lambda_{j-2}x_{j-2}(1-x_{j-1}) ,
\end{align*}
suggesting that site~$j-1$ will also fill  up. 
In this way, a traffic jam of particles is formed ``behind'' the bottleneck  rate~$\lambda_j$.

Note that if~$\lambda_j=0$ for some index~$j$ then the RFM splits into two separate chains, so we always
 assume that~$\lambda_j>0$ for all~$j \in\{0,\dots,n\}$. 

The asymptotic behavior of the RFM has been analyzed using tools from
 contraction theory~\cite{RFM_entrain}, the theory of cooperative dynamical systems~\cite{RFM_stability}, continued fractions and  Perron-Frobenius theory~\cite{RFM_sense}. 
We briefly review some of these results that are required later on.

\subsection{Dynamical Properties of the RFM} \label{sec:dyn}

Let~$x(t,a)$ denote the solution of the RFM 
at time~$t \ge 0$
   for the initial condition~$x(0)=a$. Since the  state-variables correspond to normalized occupancy levels, we always assume that~$a$ belongs to the  closed $n$-dimensional unit cube:
\[
          [0,1]^n:=\{x \in \R^n: x_i \in [0,1] ,\; i=1,\dots,n\}.
\]
Let~$(0,1)^n$ denote the interior of~$ [0,1]^n$.
It was shown in \cite{RFM_stability} (see also~\cite{RFM_entrain})
that there exists a unique~$e=e(\lambda_0,\dots,\lambda_n)\in(0,1)^n$ such that 
for any $a \in  [0,1]^n $  the solution   satisfies~$x(t,a)\in (0,1)^n$ for all~$t >  0$ and  
\[
\lim_{t\to\infty}x(t,a)=e. 
\]
In other words,   every state-variable 
 remains well-defined in the sense that it always takes values in~$[0,1]$,
and the state converges to a  unique steady-state that depends on the~$\lambda_i$s, but not on the initial condition. 
\updt{At the steady-state, the flows into and out of each site are equal, and thus the density in the site remains constant.}
The rate of convergence to the steady-state~$e$ is exponential~\cite{cast_book}.
Note that the  production rate~${R}(t)=\lambda_n x_n(t)$ converges to the steady-state  value~$ {R} :=\lambda_n e_n$, as $t\to\infty$. 

At the steady-state, the left hand-side of~\eqref{eq:rfm_all} is zero, and this gives
\be\label{eq:poyr}  \lambda_i e_i(1-e_{i+1}) =  {R} ,\quad i=0,1,\dots,n,
\ee
		where we define~$e_0 :=1$ and $e_{n+1} :=0$.
In other words, at the steady-state the flow into and out of each site are equal to~${R}$. 

Solving the set of non-linear equations in~\eqref{eq:poyr}  is not trivial. Fortunately, there exists
a better representation of the mapping
 from the rates~$\lambda_0,\dots,\lambda_n$ to the steady-state~$e_1,\dots,e_n$. Let~$\R^k_{>0}$
denote the set of~$k$-dimensional vectors with all entries positive.
Define the $(n+2)\times(n+2)$ tridiagonal   matrix
\be\label{eq:bmatrox}
                 {T}_n := \begin{bmatrix}
 0 &  \lambda_0^{-1/2}   & 0  & \dots &0&0 \\
\lambda_0^{-1/2} & 0  & \lambda_1^{-1/2}     & \dots &0&0 \\
 0& \lambda_1^{-1/2} & 0   & \dots &0&0 \\
 & &&\vdots \\
 0& 0 & 0 & \dots & 0& \lambda_{n }^{-1/2}     \\
 0& 0 & 0 & \dots  & \lambda_{n }^{-1/2}  & 0
 \end{bmatrix}.
\ee
This is a   symmetric matrix, so
 all its eigenvalues are real. 
Since every entry of~$ {T}_n$ is  non-negative  and~$ {T}_n$ is irreducible, it admits a simple 
 maximal eigenvalue~$\sigma>0$ (called the Perron eigenvalue or Perron root of~$ {T}_n$),
 and a corresponding eigenvector~$\zeta\in\R^{n+2}_{>0}$ 
(the Perron eigenvector) that is unique (up to scaling)~\cite{matrx_ana}.

Given  an RFM with dimension $n$ and rates $\LMD$,
	let~$ {T}_n$ be the matrix defined in~\eqref{eq:bmatrox}. 
	It was shown in~\cite{rfm_concave} that then
\begin{align}\label{eq:spect_rep}
 {R}  =\sigma^{-2}  \text{ and }
e_i =\lambda_i^{-1/2}\sigma^{-1}\frac{\zeta_{i+2}}{\zeta_{i+1}}, \quad i=1,\dots,n.
\end{align}
In other words,   the steady-state density and production rate in the RFM can be directly 
obtained from the spectral properties of~${T}_n$. In particular, this makes it possible to
determine~${R}$ and~$e$ even for very large chains
using     efficient and numerically stable algorithms for computing
 the Perron eigenvalue  and eigenvector  of a tridiagonal matrix.

The spectral representation has several  useful theoretical  implications. 
It implies that
that~$ {R} ={R} (\LMD)$ is a \emph{strictly concave function} on $\R^{n+1}_{>0}$.  \updt{Thus,  
the problem of maximizing~$ {R} $ under an upper bound on the sum of the rates always admits a unique solution~\cite{rfm_concave}.}

Also, the spectral representation  implies that the sensitivity of  the steady-state w.r.t. a perturbation in the rates
becomes an eigenvalue sensitivity problem.  
Known results on the sensitivity of the Perron root \updt{\cite{eigen_sense}} imply that
\begin{align}\label{eq:spect_rep_sens}
\frac{\partial }{\partial \lambda_i}  {R} &=\frac{2}{\sigma^3 \lambda_i^{3/2} \zeta'\zeta} \zeta_{i+1} \zeta_{i+2} , \quad i=0,\dots,n,
\end{align}
where~$\zeta'$ denotes  the transpose of the vector~$\zeta$. 
It follows in particular  that~$\frac{\partial }{\partial \lambda_i}  {R} >0$ for all $i$, that is,
an increase in any of the transition rates   yields an increase in the steady-state production rate~\cite{RFM_sense}. 
 
The RFM has been used to analyze various properties of translation.
These include mRNA circularization and ribosome cycling~\cite{RFM_feedback},
maximizing  the steady-state production rate under a constraint on the rates~\cite{rfm_concave,RFM_r_max_density},
optimal down regulation of translation~\cite{zarai2017optimal}, and
the effect of ribosome  drop off  on the production rate~\cite{zarai2017deterministic}.
More recent work focused on coupled networks of mRNA molecules.
The coupling may be due to  competition for shared resources like the finite pool 
of free ribosomes~\cite{Raveh2016,splitting}, or due to the effect of the proteins produced 
on the promoters   of other mRNAs~\cite{nani}.
Several variations and
generalizations of the RFM have also  been suggested and analyzed~\cite{RFMR,russ1,10.1371/journal.pone.0182074,zarai2017deterministic,Zarai20170128,EYAL_RFMD1}.

Several studies  compared  predictions of the RFM with  biological   measurements.
For example,
protein levels and ribosome densities in translation~\cite{reuveni},
 and RNAP densities in transcription~\cite{Edri2014}.
The results demonstrate high correlation
  between   gene expression measurements and the RFM predictions.

  
All previous works on the RFM   assumed that the transition rates~$\lambda_i$ are
deterministic.  Here, we analyze for the first time the case where
the  rates are random variables.  	This  may  model for example 
	the parallel translation  of copies of the same mRNA molecule in different locations inside 
	the cell. 
	The variance of factors like  tRNA abundance 
	in  these different locations   implies that
	each mRNA is translated with different rates.
	It is natural to model this variability using tools from probability theory. 
 For example,  Ref.~\cite{dana2014effect}
 showed that the distribution of read counts related to a codon in ribo-seq experiments can be approximated using an  exponentially modified Gaussian. 

\updt{ 
Our results analyze the  average steady-state production rate  given the random transition rates. Note that this provides a  global   picture 
of   protein production in the cell, rather than the local production in any single chain.
For example,
   when ``drawing'' the rates from a  given distribution,  one rate may turn  out to be much smaller than the others and  this will generate a traffic jam in the corresponding chain.  
However, our analysis does not consider any specific  chain, but the average steady-state production rate on all the chains drawn according to the distribution of the i.i.d. rates.
}

 The following section describes our main results
on translation with random  rates.
\section{Main Results }\label{sec:main}

Assume that the~RFM  rates are not constant, but rather 
 are  random variables  with
some known distribution   supported over~$\R_{\geq \delta}\ := \{x\in\R : x\geq \delta \} $, where~$\delta>0$.
What will   the statistical properties of the resulting  protein production rate be?
 In the context of the spectral representation given in~\eqref{eq:bmatrox},
 this amounts to the following question: given the distributions of the
random variables~$\{\lambda_i\}_{i=0}^n$, what are the statistical properties
of the maximal eigenvalue~$\sigma$ of the random  matrix~$\s{T}_n$?

Recall that a random variable $\s{X}$ is called \emph{essentially bounded}
 if there exists~$0\leq {b}<\infty$ such that~$\pr\pp{\abs{\s{X}}\leq b}=1$, and then the~$L_\infty$ norm of~$\s{X}$ is
\[
 \|\s{X}\|_\infty : = \inf_{ {b}\geq0}  \ppp{ \pr\pp{\abs{\s{X}}\leq   {b}}=1}.
\]  
\updt{Roughly speaking, this is the maximal value that~$\s{X}$ can typically attain. 
}
Clearly,   bounded random variables is    the relevant case in  any biological  model.
In particular, if~$\s{X}$ is supported over~$\R_{\geq \delta}$, with~$\delta>0$,
then the~random variable defined by~$\s{W} := \s{X}^{-1/2}$ is essentially bounded and~$||\s{W} ||_\infty \leq  \delta^{-1/2} $. 

We can now state our main results. To increase readability, all 
the proofs are placed in the final section of this paper.
\updt{To emphasize that now the production  rate is a random variable, and that it depends on the length of the chain, from hereon we  use~$\s{R}_n$ to denote the production rate in the~$n$-site RFM. }

\begin{Theorem}\label{thm:main}
Suppose that every rate~$\lambda_0,\dots,\lambda_n$ in the RFM
is drawn independently according to the distribution of a  random variable~$ \s{X}$ that is supported on~$\mathbb{R}_{\geq \delta}$, with~$\delta>0$. 
Then as~$n\to \infty$,
the maximal eigenvalue of the matrix~$\s{T}_n$ converges
 to~$2 ||\s{X}^{-1/2}||_\infty $
with probability one,
 and  the steady-state production rate~$\s{R}_n$ in the~RFM converges to
\be\label{eq:spor}
 (2 ||\s{X}^{-1/2}||_\infty)^{-2}, 
\ee
with probability one. 
\end{Theorem}
This result may   explain how proper functioning is maintained in spite of
 significant  variability in the rates:   the steady-state production rate always converges to the value
in~\eqref{eq:spor}, that depends only on~$||\s{X}^{-1/2}||_\infty$. 
This also  implies a form of \updt{universality}  with respect to the   noises and uncertainties:
the exact details of the distribution of~$\s{X}$ are not relevant, 
but only the single value~$||\s{X}^{-1/2}||_\infty$.

\begin{figure}[t]
\begin{center}
\includegraphics[scale=0.65]{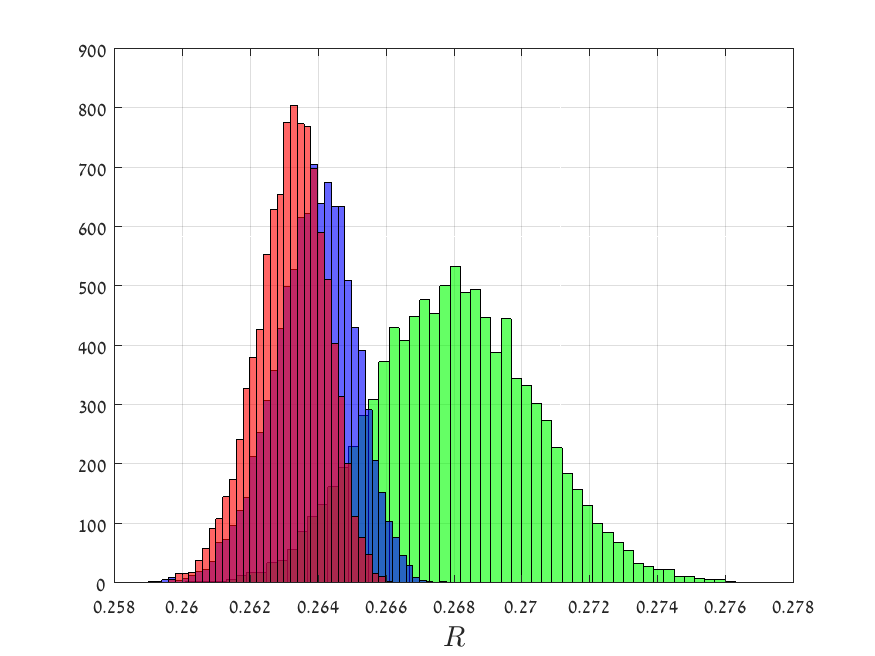}
\caption{Histograms of $10,000$ $\s{R}_n$ values in Example~\ref{exa:chi1} for~$n=50$~(green), $n=500$~(blue), and ~$n=1000$~(red). The theory predicts
that as~$n\to\infty$,
 $\s{R}_n$ converges to~$1/4$ with probability one.}  \label{fig:hist_10000_4}
\end{center}
\end{figure}

In general, the convergence to the values in Theorem~\ref{thm:main} as~$n$ increases is slow,
and computer  simulations  may require~$n$ values that exhaust the computer's memory 
 before we are close to the theoretical values. 
The next example demonstrates a case where the convergence is relatively fast. 
\begin{Example}\label{exa:chi1}
Recall that the probability density function of the 
half-normal distribution with parameters~$(\mu,\sigma)$ is 
\[
					f(x)= \begin{cases} 
					      \sqrt{ \frac{2}{\pi\sigma^2}}  \exp( -\frac{1}{2} (\frac{x-\mu}{\sigma} )^2  ),   &  x\geq  \mu,\\  
								0, & \text{otherwise}.
								\end{cases}
\]
\updt{This may be interpreted as a kind of normal distribution, but with support over~$[\mu,\infty)$ only}. Suppose that~$\s{X}$ has this distribution  
with parameters~$( \mu=1,\sigma= 0.1 )$. \updt{
Note that~$\s{X}^{-1/2}$ has support~$(0,1]$, so~$||\s{X}^{-1/2}||_\infty =1$.}
In this case, Thm.~\ref{thm:main} implies that~$\s{R}_n$ converges with probability one to~$1/4$ as~$n$ goes to infinity. 
For~$n\in\{50,500,1000\}$, we numerically computed~$\s{R}_n$ using the spectral representation 
for~$10,000$ random matrices. Fig.~\ref{fig:hist_10000_4} depicts a histogram of the results.
 It may be seen that as~$n$ increases the histogram becomes ``sharper''
and its center  
  converges towards~$1/4$, as expected.
\end{Example}

Theorem~\ref{thm:main} does not provide  any information on the rate of convergence to 
the limiting value of~$\s{R}_n$.
This is important as in practice~$n$ is always finite.
The next result  addresses this issue.
For~$\epsilon>0$,
let
\[
a(\epsilon):= \pr\p{ \s{X}^{-1/2}  \geq  \|\s{X}^{-1/2}\|_\infty-\epsilon}.
\]
Note that~$a(\epsilon) \in (0,1)$.
\updt{Intuitively speaking, $a(\epsilon)$ 
is the probability that~$\s{X}^{-1/2}$ falls in the range~$[\|\s{X}^{-1/2}\|_\infty-\epsilon,\|\s{X}^{-1/2}\|_\infty ]$. 

}
\begin{Theorem}\label{thm:main_with_more}
Suppose that every rate~$\lambda_0,\dots,\lambda_n$ in the RFM
is drawn independently according to the distribution of an random variable~$ \s{X}$ that is supported on~$\mathbb{R}_{\geq \delta}$, with~$\delta>0$. 
Pick  two sequences of positive 
integers~$n_1<n_2<\dots$ and~$k_1<k_2<\dots$, with~$k_i<n_i$ for all~$i$,
and a decreasing sequence of positive scalars~$\epsilon_i$, with~$\epsilon_i \to 0$.
Then for any~$i$ 
the steady-state production rate~$\s{R}_{n_i} $ in an RFM with~$n_i$
sites satisfies  
\begin{align}\label{eq:potr}
    (2\|\s{X^{-1/2}}\|_\infty)^{-2}\leq \s{R}_{n_i}&\leq (2\|\s{X^{-1/2}}\|_\infty)^{-2}
		\left( 1+O(\epsilon_i+k_i^{-2}) \right ),
\end{align}
with probability at least 
\begin{align} 
  1-\exp\left ( - { \Bigl \lfloor  \frac{n_i-1}{k_i} \Bigl \rfloor} ( a(\epsilon_i))^{k_i } \right )  .
\end{align}
\end{Theorem}

Note that if we choose the sequences  
 such that 
\begin{align}\label{Eq:infc}
     \frac{n_i}{k_i} (a(\epsilon_i))^{k_i} \to \infty,
\end{align}
and  take~$i \to\infty$ then Theorem~\ref{thm:main_with_more} yields  
  Theorem~\ref{thm:main}. Yet, we  state and prove 
both results separately in the interest of  readability. 

\begin{Example}\label{exa:chi2}
Suppose that~$\s{X} $ has a uniform distribution over an interval~$[\delta,\gamma]$
 with~$0< \delta<\gamma$. From here on we assume for simplicity that~$\delta=1$ and~$\gamma=2$.
Then for any~$\epsilon >0$ sufficiently small, we have
\begin{align*}
a(\epsilon)&= \pr\p{ \s{X}^{-1/2} \geq 1-\epsilon  }\\  
          &=  \pr\p{ \s{X} \leq (1-\epsilon)^{-2}  }\\
          &=2\epsilon+o(\epsilon).
\end{align*}
Fix~$ d\in(0,1)$ and
take~$\epsilon_i = n_i^{(d-1)/k_i}$. 
 Then the condition in~\eqref{Eq:infc} becomes 
\[
 \frac{n_i^d} {k_i}  \to \infty
 \]
 and this will hold if~$k_i$ does not increase too quickly. We can write~$\epsilon_i$ as
\[
\epsilon_i=\exp(  (d-1)\log(n_i)/k_i ),
\]
so to guarantee that~$\epsilon_i  \to 0 $, 
we take~$k_i=(\log(n_i))^c$, with~$c\in(0,1)$, and then~\eqref{Eq:infc} indeed holds.
Theorem~\ref{thm:main_with_more}  implies that
\begin{align*}
   & (2\|\s{X^{-1/2}}\|_\infty)^{-2}\leq \s{R}_{n_i} \leq (2\|\s{X^{-1/2}}\|_\infty)^{-2}
   \left( 1+O( \max\{\exp(  (d-1) (\log(n_i) )^{1-c} )    , (\log(n_i))^{-2c}  \}    ) \right ),
\end{align*}
with probability at least 
\begin{align} 
  1-\exp\left (  \frac{-n_i^d}{(\log(n_i))^c} \right )  .
\end{align}
\end{Example}

\begin{Example}
As   in Example~\ref{exa:chi1},
consider the case where $\s{X}$ is half-normal with parameters~$(\mu,\sigma )$, where~$\mu>0$.
Then~$\|\s{X}^{-1/2}\|_\infty = \mu^{-1/2}$, so
\begin{align*}
a(\epsilon)&= \pr\p{ \s{X} ^{-1/2} \geq  \mu^{-1/2}-\epsilon} \\
&= \pr\p{\s{X}\leq z },
\end{align*}
where~$z := (\mu^{-1/2}-\epsilon)^{-2}$. Thus,
\begin{align*}
a(\epsilon)
& = \sqrt{\frac{2}{\pi\sigma^2}}\int_{\mu}^{z}e^{-\frac{(x-\mu)^2}{2\sigma^2}}\mathrm{d}x\\
& = \frac{2}{ \sqrt{\pi}}\int_{0}^{\frac{ z-\mu}{\sqrt{2\sigma^2}}}e^{-x^2}\mathrm{d}x.
\end{align*}
It is not difficult to show that this implies that
\begin{align}
a(\epsilon) 
=  c(\mu,\sigma) \epsilon+o(\epsilon)  , 
\end{align}
where $c(\mu,\sigma) : =    2 \sqrt{ \frac{2}{\pi \sigma^2}} \mu^{3/2} $.
To satisfy~\eqref{Eq:infc}, fix~$p\in (0,1) $ and choose
 $\epsilon_i$ such that $(c \epsilon_i)^{k_i} 
 = n_i^{p-1}$. This implies that 
\begin{align}
  \epsilon_i = \frac{1}{c }  \exp\p{\frac{p-1}{k_i}\log(n_i)}.
\end{align}
Now, pick~$q\in(0,1)$ and take~$k_i = (\log( n_i))^q$. Then~\eqref{Eq:infc} holds, and
\begin{align}
\epsilon_i = \frac{1}{c } \exp\p{ (p-1) ( \log(n_i) ) ^{1-q}}.
\end{align}
Theorem~\ref{thm:main_with_more} implies that for any~$p,q\in(0,1)$, we have
\begin{align*}
 \frac{\mu}{4}\leq \s{R}_{n_i} \leq \frac{\mu}{4} +O( \max\{   \frac{1}{c } \exp\p{ (p-1) ( \log(n_i) ) ^{1-q}}, (\log(n_i))^{-2q}    \} ),
\end{align*}
with probability at least 
\[
1-\exp\left (  \frac{-n_i^p}{  (\log(n_i))^q }        \right ).
\]
\updt{This shows that~$\s{R}_{n_i}$ ``is close'' to~$\mu/4$, and provides an explicit expression for the rate of convergence to~$\mu/4$.}
\end{Example}

\updt{
\section{Generalizations}\label{sec:gen} }
\updt{
The assumption that all the rates are i.i.d.  random variables allows to derive the general  theoretical results in Theorems~\ref{thm:main}
and~\ref{thm:main_with_more} above.
However, this assumption is restrictive. In this section, we describe several cases  where we allow more relaxed  assumptions on these rates. Our first generalization considers the case where the random variables might be non-identical, but  all share the same support. In the second generalization, we allow an increasing   (but small compared to $n$) number of random variables to have a different support from the majority of the other random variables. In these two cases we show that the production rate converges to the same value as in Theorem~\ref{thm:main}. 
We then turn to  investigate the most general case, where the rates are arbitrary but bounded, and in this case provide lower and upper bounds on the production rate.}

Analysis of  the proofs of Theorems~\ref{thm:main} and~\ref{thm:main_with_more} shows that our results remain valid even if each rate~$\lambda_i$ is drawn from the distribution of $\s{X}_i$, which are not necessarily identically distributed, but are all independent, supported on the positive semi-axis, and satisfy
\be\label{eq:sames}
||\s{X}_0^{-1/2}||_\infty=\dots=||\s{X}_{n}^{-1/2}||_\infty,
\ee
namely, they all have the same bound. \updt{The next example demonstrates this.} 
 
\begin{figure}[t]
\begin{center}
\includegraphics[scale=0.65]{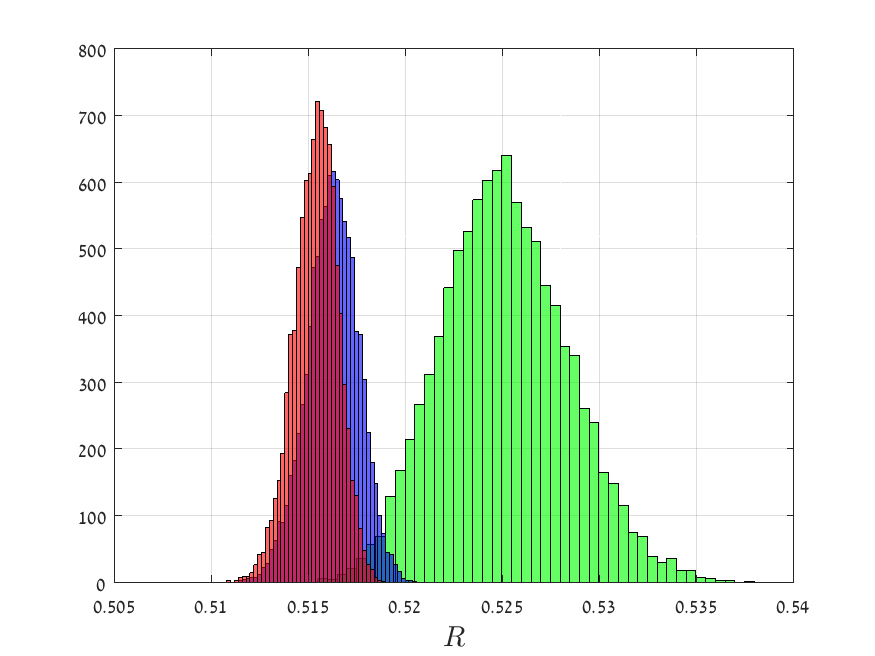}
\caption{Histograms of $10,000$ $\s{R}_n$ values in Example~\ref{exa:chi4} for~$n=50$~(green), $n=500$~(blue), and ~$n=1000$~(red). The theory predicts
that as~$n\to\infty$,
 $\s{R}_n$ converges to~$1/2$ with probability one.}  \label{fig:hist_10000_4_added}
\end{center}
\end{figure}

\begin{Example}\label{exa:chi4}
\updt{
Consider~$n+1$ independent random variables 
with~$\s{X}_{0},\s{X}_{1},\dots,\s{X}_{\frac{n-1}{2}}$  distributed  according to the 
half-normal distribution
with parameters~$( \mu=2,\sigma= 0.1 )$, and~$\s{X}_{\frac{n-1}{2}+1},\dots, \s{X}_{n}$  distributed  according to the uniform distribution on~$[2,3]$. 
Note that~$||\s{X_i}^{-1/2}||_\infty =2^{-1/2}=1 / \sqrt{2}$, for all~$i=0,1,\dots,n$. 
Thus,   our theory predicts that  in this case~$\s{R}_n$ converges with probability one to~$ (2/\sqrt{2})^{- 2} = 1/2$ as~$n$ goes to infinity. 
For~$n\in\{50,500,1000\}$, we numerically computed~$\s{R}_n$ using the spectral representation 
for~$10,000$ random matrices. Fig.~\ref{fig:hist_10000_4_added} depicts a histogram of the results.
 It may be seen that as~$n$ increases the histogram becomes ``sharper''
and its center  
  converges towards~$1/2$, as expected.
} 
\end{Example}

\updt{
Our second generalization considers the case where among the~$n+1$  random rates  there are~$d$ rates drawn from the distributions of the random variables~$\s{Y}_1,\dots, \s{Y}_d$, that might have some different distributions; they do not have to satisfy the uniform support condition in \eqref{eq:sames}, and they might be dependent. Here~$d=d(n)$ is an integer that is allowed to grow with~$n$, but at a slower rate than~$n$. We assume that the rates modeled by these random variable are larger those rates modeled by the other $n+1-d$ random variables (see  \eqref{eq:ybound} below).}
 \begin{Theorem}\label{thm:event}
Let $d=d(n)>0$ be an integer such that $\lim_{n\to\infty} 
\frac{d(n)}{n} =0$.
Let $\{\s{X}_i\}_{i=0}^{n-d}$ be a set of~$(n+1-d)$ independent random variables, supported on $\mathbb{R}_{\geq\delta}$, with $\delta>0$, and satisfying
\[
||\s{X}_0^{-1/2}||_\infty=\dots=||\s{X}_{n-d}^{-1/2}||_\infty.
\]
Also, let $\{\s{Y}_i\}_{i=1}^d$ be a set of $d$ random variables supported on the positive semi-axis, and satisfy 
\be\label{eq:ybound}
||\s{Y}_j^{-1/2}||_\infty \leq \delta^{-1/2}, \quad j=1,\dots,d.
\ee
Fix $\epsilon>0$ and a positive integer~$k$. Denote the concatenation of $\{\s{Y}_i\}_{i=1}^d$ and $\{\s{X}_i\}_{i=0}^{n-d }$ by $\s{Z}$, namely, $\s{Z} := (\s{Y}_1,\s{Y}_2,\ldots,\s{Y}_d,\s{X}_0,\s{X}_1,\ldots,\s{X}_{n-d  })$. 
Let~$\mathcal S^{n+1}$ denote the set of permutations on~$\{1,\dots,n+1\}$.
Fix  a permutation~$\pi \in \mathcal S^{n+1}$, 
 and let $\s{Z}^\pi \triangleq  \pi\circ\s{Z}$. 
Suppose that every rate~$\lambda_i$ in the~RFM
is drawn independently according to the distribution of the random variable  in~$\s{Z}^\pi_i$. 
Then as~$n\to \infty$,
  the steady-state production rate~$\s{R}_n$ in the~RFM converges to
\be\label{eq:spor2}
 (2 ||\s{X}_0^{-1/2}||_\infty)^{-2}, 
\ee
with probability one. 
\end{Theorem}
In other words, even in the presence of  the ``interfering''~$\s{Y}_i$'s
the theoretical   result remains unchanged.

\begin{figure}[t]
\begin{center}
\includegraphics[scale=0.65]{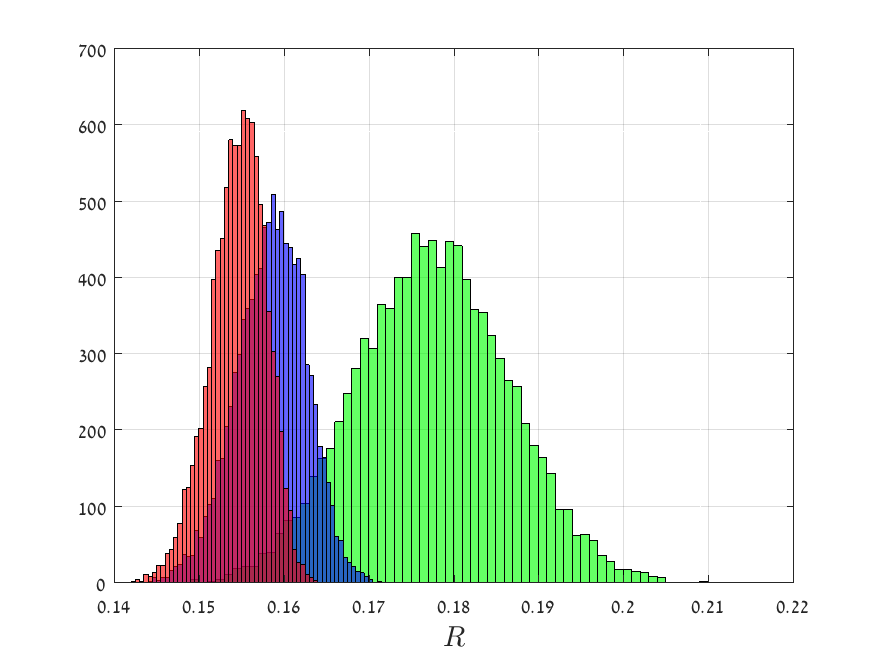}
\caption{Histograms of $10,000$ $\s{R}_n$ values in Example~\ref{exa:withdn} for~$n=50$~(green), $n=500$~(blue), and ~$n=1500$~(red). The theory predicts
that as~$n\to\infty$,
 $\s{R}_n$ converges to~$1/8$ with probability one.}  \label{fig:hist_10000_4_yis}
\end{center}
\end{figure}

 The next example demonstrates Theorem~\ref{thm:event}.
\begin{Example}\label{exa:withdn}
Consider the case where~$d(n)=\sqrt{n}$. 
Let~$\s{X}_0,\dots,\s{X}_{n-d}$
be i.i.d. random variables distributed according to the uniform 
distribution on~$[1/2,1]$,
and let 
Let~$\s{Y}_1,\dots,\s{Y}_{d}$
be i.i.d.  random variables distributed according to the uniform 
distribution on~$[15,20]$.  
We draw the rates according to the vector~$\s{Z}^{\pi}$,
with~$\pi$ a random permutation 
(implemented using the Matlab command \emph{randperm}). 
Our theory predicts that  in this case~$\s{R}_n$ converges with probability one to~$ 
(2||\s{X}_i^{-1/2}||_\infty)^{-2} =(2 \sqrt{2})^{-2}=1/8 $ as~$n$ goes to infinity. 
For~$n\in\{50,500,1500\}$, we numerically computed~$\s{R}_n$ using the spectral representation 
for~$10,000$ random matrices. Fig.~\ref{fig:hist_10000_4_yis} depicts a histogram of the results.  It can be seen that the~$\s{R}_n$ converges with probability one to a limiting value, despite the ``interfering'' $\s{Y}_i$ random variables.  
\end{Example}

\updt{
Our last and most general result considers the case where the random variables are arbitrary but bounded. In particular, they do not necessarily have to be independent or identical. We use the notation~$\mathcal{I}_k^{p}$ to denote the set of all possible~$k$ consecutive integers from the set~$\{1,2,\dots,p\}$. For example, 
\[
\mathcal{I}_2^3 = \{ \{1,2\}, \{1,3\}, \{2,3\}  \}.
\]
}
\updt{
\begin{Theorem}\label{thm:event2}
Suppose that every rate~$\lambda_i$    in the RFM
is drawn according to the distribution of a random variable~$ \s{X}_i$ that is supported on~$\mathbb{R}_{\geq \delta_i}$, with~$\delta_i>0$, for $0\leq i\leq n$. Then  the steady-state production rate~$\s{R}_n$ in the~$n$-site RFM satisfies  
\begin{align}
    \left[\max_{i=1,\ldots,n}\s{X}_{i-1}^{-1/2}+\s{X}_i^{-1/2}\right]^{-2}\leq \s{R}_n&\leq \left[2\max_{1\leq k\leq n+1}\cos\left(\frac{\pi}{k+2}\right)\max_{\s{I}_k\in\mathcal{I}_k^{n+1}}\min_{i\in\s{I}_k}\s{X}_i^{-1/2}\right]^{-2},\label{eqn:Rgeneral}
\end{align}
with probability one.
\end{Theorem}
}
\updt{Contrary to our previous analytical  results,  in this case the steady-state production rate will not necessarily converge to a deterministic value, but rather we show that it is  bounded above and below by two random quantities. However, it can be shown that  when the random variables are i.i.d. then
 both bounds converge to~$(2||\s{X}_0^{-1/2}||_\infty)^{-2}$ as $n\to\infty$, and in this sense 
 the bounds in Theorem~\ref{thm:event2} are tight.}

\section{Discussion}
Cellular  systems are inherently noisy, and it is natural to speculate  that
they were  optimized by evolution to function properly, or even take advantage,
  of stochastic  fluctuations.  

Many studies analyzed the  
fluctuations in protein production  
due to   both extrinsic and intrinsic noise (see, e.g.~\cite{sum_noise,noise_single,trade-off2019,arkin_noise,rati2020,Zarai2017}). 
Here, we derived  a new  approach, based on random matrix theory,  for analyzing  the average    protein production rate  from multiple  copies of the same mRNA that are 
affected 
by
variations in the translation rates due, for example, 
to   the  different spatial location of these   mRNAs inside the cell. Our approach can also deal with experimental noise.  

\updt{
Our results have both a theoretical and  a practical value. We show that  given one parameter value~$\delta$ from the i.i.d. distribution  allows to  determine the steady-state average production rate. 
The production rate  is thus agnostic to many other details underlying the distribution e.g. it's mean, variance, etc. This may explain how steady-state production is maintained despite the considerable stochasticity  in the cell. 
 This theoretical result holds regardless of whether one can actually determine the value~$\delta$ or not.
}

\updt{
Our approach can also deal with   phenomena that is not directly captured by the~RFM, if its affects can be modeled 
as a stochastic perturbation of the transition rates. 
Examples may include 
experimental noise, methylation, and 
interaction with miRNA.
In particular,  methylation affects one nucleotide/codon, and miRNA affects a sequence of   up to 7 codons.
}

It is important to note  that our results hold
for many possible distributions of the translation rates. For example, it was suggested that decoding rates distributions are similar an exponential modified Gaussian   or log normal  distribution~\cite{dana2014effect,Dana2014A}.

\updt{Currently, it is challenging   to estimate the distribution of transition rates (and thus the bound on the  support~$\delta$).
Indeed, approaches such as ribo-seq provide  averages over all mRNA molecules and all cells in a certain population/sample.
It is
 also difficult to estimate the protein translation rate. Usually, the measured quantity is 
protein level, but this  depends not only on translation, but also on the rate of transcription, and mRNA and protein dilution and decay~\cite{trade-off2019}. 
  Thus, in this respect, the theory in the paper precedes biological measurement capabilities.}
Our results however may indicate  general  principles that can be tested experimentally. For example,  the analysis suggests that as the length of the mRNA increases,  while keeping all the statistical properties  that determine
initiation rate and codon usage identical, the translation rate becomes more uniform.

The RFM, just like TASEP, is a phenomenological model for the flow of interacting particles and 
thus can be used to model and analyze phenomena like the flow of packets in communication networks~\cite{rfm_wireless}, the  transfer of a
phosphate group   through a serial chain of proteins during phosphorelay~\cite{EYAL_RFMD1}, and more.
The RFM is also closely related  to a mathematical model for 
a disordered
linear chain of masses, each coupled to its two nearest neighbors by elastic springs~\cite{rfm_dyson}, that was originally analyzed in the seminal work of Dyson~\cite{dyson53}. 
 In many of these applications it is natural to assume that the rates are 
subject to uncertainties  or fluctuations and model them as random variables. Then the results here can be immediately applied. 

We believe that the approach described here can be generalized  to other     models of intra-cellular phenomena derived from the RFM~\cite{Zarai2018SB,EYAL_RFMD1},
and thus   for analyzing  additional aspects of translation and gene expression.

\section{Proofs}\label{sec:proofs}
The proofs of our main results are  based on analyzing the \updt{spectral properties of the} matrix~$\s{T}_n$ in~\eqref{eq:bmatrox} 
when the~$\lambda_i$s are i.i.d. random variables. 	The problem that
we study here is a classical problem in random matrix theory~\cite{random_mat_silver}, yet the matrix~$\s{T}_n$   is somewhat different from the standard matrices analyzed  using the existing theory (e.g. the Wigner  matrix). Hence, we provide   a self-contained analysis based on combining  probabilistic arguments with the Perron-Frobenius theory of matrices with non-negative entries (see e.g.~\cite[Ch.~8]{matrx_ana}).

\subsection{Proof of Theorem~\ref{thm:main}}
Recall that the rates $\{\lambda_i\}_{i=0}^{n}$ are drawn independently according to the distribution of a random variable $\s{X}$ that is supported on $\mathbb{R}_{\geq\delta}$, with $\delta>0$. For simplicity of notation, let~$\s{W}_i := \lambda_i^{-1/2}$,   $i\in\{0,1,\ldots,n\}$, and   note that $\{\s{W}_i\}_{i=0}^{n}$ are essentially bounded, i.i.d., and each random variable $\s{W}_i$ follows the same distribution of $\s{X}^{-1/2}$. In particular,   $\s{W}_0\equiv\s{X}^{-1/2}$. With this definition,   \eqref{eq:bmatrox} can be written as:
\begin{align}
\s{T}_n :=  \begin{pmatrix}0 & \s{W}_0 &  & & \\ \s{W}_0 & 0 & \s{W}_1 &  & \\  & \s{W}_1 & 0 & \ddots & \\ &  & \ddots & & \s{W}_{n} \\ & & & \s{W}_{n} & 0\end{pmatrix}.\label{eqn:TriMatrix}
\end{align}
Therefore, $\s{T}_n $ is an $(n+2)\times(n+2)$ symmetric tridiagonal matrix, with zeros on its main diagonal, and bounded positive random variables $\{\s{W}_i\}_{i=0}^{n}$ on the super- and sub-diagonals.
 
Since~$\s{T}_n$ is symmetric, componentwise non-negative, and irreducible, it admits  
a simple maximal eigenvalue denoted~$\lambda_{\max}(\s{T}_n)$, and~$\lambda_{\max}(\s{T}_n) >0$. 
Our goal is to  understand the asymptotic  behavior of $\lambda_{ \max}(\s{T}_n )$, as~$n\to\infty$. 
We begin with an auxiliary result that will be used later on.

\begin{Proposition}\label{prop:event}
Suppose that the random variables $\{\s{W}_i\}_{i=0}^{n}$ are i.i.d. and essentially bounded.  Fix~$\epsilon>0$ and an integer~$1\leq k\leq n+1$.
Let $\cal K$ denote the event:  there exists an
 index~$0\leq\ell \leq n-k+1$ such that~$ \s{W}_{\ell },\dots,\s{W}_{\ell+k-1 }\geq \|\s{W}_0\|_\infty-\epsilon$.
Then as~$n\rightarrow \infty$ the probability of $\cal K$ converges to one. 
\end{Proposition}

\updt{In other words, as~$n\to \infty$ the probability of finding~$k$  consecutive random variables whose value is at least~$\|\s{W}_0\|_\infty-\epsilon$  goes to one.}

\begin{Proof*} 
Fix $\epsilon>0$    and a positive integer~$k$. Let~$s : = \|\s{W}_0\|_\infty-\epsilon$.
For any~$j\in\{0,\dots,n-k+1\}$, let~${\cal K} (j)$  denote the event:
  $ \s{W}_{j },\dots,\s{W}_{j+k -1}\geq s$.
Then
\begin{align*}
							\pr\p{{\cal K}}			&\geq 	\pr\p{ {\cal K}(1) \cup   {\cal K}(k+1) \cup{\cal K}(2k+1) \cup \dots \cup {\cal K}(pk+1)  },
\end{align*}
where~$p$ is the largest integer such that~$(p+1)k\leq n$. Since the~$\s{W}_i$s are i.i.d., 
\begin{align*}
\pr\p{{\cal K}}			&\geq 	 1-(1-\pr\p{ {\cal K}(1) } )^{p+1}\\
& = 1-(1-(\pr\p{ \s{W}_0 \geq s })^k )^{p+1}.
\end{align*}
The probability~$\pr \p{ \s{W}_0 \geq s }$ is positive, and
when~$n\to \infty$, we have~$p\to \infty$, so~$\pr\p{{\cal K}}	\to 1$.~\hfill{$\square$}
\end{Proof*}

The next result invokes Proposition~\ref{prop:event} to provide a tight asymptotic lower bound on the maximal eigenvalue of~$\s{T}_n$. 
\begin{Proposition}\label{prop:upper}
Suppose that the random variables $\{\s{W}_i\}_{i=0}^{n}$ are i.i.d. and essentially bounded. Fix~$\epsilon>0$  and an integer~$1\leq k\leq n+1$. Then the probability
\begin{align}
\pr\p{ \lambda_{\max}(\s{T}_n) \geq2 (\|\s{W}_0\|_\infty-\epsilon) \cos{\frac{\pi}{k+2}} },
\end{align}
goes to one as~$n\rightarrow \infty$.
\end{Proposition}

\begin{Proof*}
Let~$s := \|\s{W}_0\|_\infty-\epsilon$. Conditioned on the event~${\cal K}$, Proposition~\ref{prop:event} implies that 
there exists an index~$\ell$  such that~$ \s{W}_{\ell },\dots,\s{W}_{\ell+k-1 }\geq s$.
Assume that~$\ell=0$ (the proof in the case~$\ell>0$ is very similar).
Let~$\s{M}_{k}$ denote the $(k+1)\times (k+1)$ symmetric tridiagonal matrix:
\begin{align}
\s{M}_{k} : =\begin{pmatrix}0 & 1 &  & & \\ 1 & 0 & 1 &  & \\  & 1 & 0 & \ddots & \\ &  & \ddots & & 1 \\ & & & 1 & 0\end{pmatrix}.
\end{align} 
Recall that the maximal eigenvalue of this matrix  
is $\lambda_{\s{max}} (\s{M}_{k }) =2 \cos{\frac{\pi}{k+2}}$ (see e.g.~\cite{eig_tridiagonal_fon}). Now, let~$ \s{P}_n$ be the matrix obtained by replacing the~$(k+1)\times (k+1)$ leading principal minor 
of~$ \s{T}_n$ by~$s \s{M}_k$. Note that~$ \s{T}_n \geq \s{P}_n$ (where the inequality  is componentwise), and thus~$\lambda_{\max} (\s{T}_n) \geq \lambda_{\max} (\s{P}_n)$. By Cauchy's interlacing theorem, the  largest eigenvalue of~$\s{P}_n$ is larger or equal to the largest eigenvalue of any of its principal minors. Thus,
\begin{align*}
\lambda_{\max} (\s{P}_n)&\geq\lambda_{\max} (s\s{M}_k)\\
& \geq 2 s \cos\p{{\frac{\pi}{k+2}}}.
\end{align*}
and this completes the proof of Proposition~\ref{prop:upper}.~\hfill{$\square$}
\end{Proof*}
  
We can now complete the proof  of Theorem~\ref{thm:main}.
Recall that if~$ {A} $ is an~$n\times n$ symmetric and componentwise non-negative matrix then (see, e.g.~\cite[Ch.~8]{matrx_ana})
\begin{align}
 \lambda_{\s{max}}( {A} ) \leq\max_{i\in\{1,\dots,n\}}\sum_{j=1}^n a_{ij}.\label{eq:micw}    
\end{align}
\updt{In other words,~$ \lambda_{\s{max}}( {A} )   $ is bounded from above by the maximum of the row sums of~$A$.}
As any row of $\s{T}_n$ has at most two nonzero elements,  \eqref{eq:micw} implies that 
\begin{align}\label{eq:uperrbnd}
\lambda_{\s{max}}(\s{T}_n) &\leq\max_{i \in \{1,\dots,n \} } ( \s{W}_{i-1}+\s{W}_{i} )\nonumber\\& \leq 
2\max_{i\in \{0,\dots,n \} }\s{W}_i, 
\end{align}
with probability one. Combining this with Proposition~\ref{prop:upper} implies that
\begin{align}\label{eq:ulbnd}
2 (||\s{W}_0||_\infty -\epsilon ) \cos\p{{\frac{\pi}{k+2}}}      \leq\lambda_{\max}(\s{T}_n)\leq 2 ||\s{W}_0||_\infty,    
\end{align}
with probability one. Since this holds for any~$\epsilon>0$ and any
integer~$k>0$, this completes the proof of Theorem~\ref{thm:main}.~\hfill{$\square$}

\subsection{Proof of Theorem~\ref{thm:main_with_more}}

Fix~$\epsilon>0$ and an integer~$1\leq k\leq n+1$. Let~$\bar a(\epsilon): =  \pr\p{\s{W}_0  \geq  \|\s{W}_0\|_\infty-\epsilon}$. The proofs of Propositions~\ref{prop:event} and~\ref{prop:upper} imply that  
\begin{align}
\lambda_{\s{max}}(\s{T}_n ) \geq2 (\|\s{W}_0\|_\infty-\epsilon) \cos{\frac{\pi}{k+2}},\label{eqn:lowerBoundLarmbda}
\end{align}
with probability $\pr(\calK)\geq1-(1-(\bar a(\epsilon)) ^k )^{\left\lfloor  \frac{n}{k} \right\rfloor}$. Fix~$b,c>0$. The trivial bound~$1-b<\exp(-b)$ implies that $1-(1-b)^{c }  >  1-\exp(-bc)$, and thus,
\begin{align}\label{eq:prcpko}
\pr(\calK)& \geq 1-(1-(\bar a(\epsilon)) ^k )^ {\left\lfloor  \frac{n}{k} \right\rfloor}\nonumber \\
& \geq 1-\exp \left ( - \left\lfloor  \frac{n}{k} \right\rfloor (\bar a(\epsilon)) ^k \right ).
\end{align}
Pick two sequences of positive  integers~$n_1<n_2<\dots$ and~$k_1<k_2<\dots$, with~$k_i<n_i$ for all~$i$,
and a decreasing sequence of positive scalars~$\epsilon_i$, with~$\epsilon_i \to 0$. Using  \eqref{eqn:lowerBoundLarmbda} we get
\begin{align*}
   ( \lambda_{\max}   (  \s{T}_{n_i} ) )^{-2}  
		& \leq \left ( 2 (\|\s{W}_0\|_\infty-\epsilon_i) \cos{\frac{\pi}{k_i+2}} \right )^{-2}\\
    &= (2\|\s{W }_0\|_\infty)^{-2} \left ( 1+\frac{\epsilon_i}{\|\s{W }_0\|_\infty}  +o(\epsilon_i) \right )  \left (\cos{\frac{\pi}{k_i+2}} \right )^{-2}\\
    &= (2\|\s{W}_0\|_\infty)^{-2}
		\left ( 1+\frac{\epsilon_i}{\|\s{W}_0\|_\infty} +o(\epsilon_i)  \right)
		\left ( 1+\frac{\pi^2}{(k_i+2)^2} + o(k_i^{-2}) \right) \\
    & = (2\|\s{W}_0\|_\infty)^{-2} \left( 1+O(\epsilon_i+k_i^{-2})\right).
\end{align*}
Combining this with the spectral representation of the steady-state in the~RFM
  completes the proof of Theorem~\ref{thm:main_with_more}.~\hfill{$\square$}

\updt{
 The proofs of Theorems~\ref{thm:event}  and~\ref{thm:event2} below are similar to the proof of Theorem~\ref{thm:main}, and so we only explain the needed  modifications in the proof of Theorem~\ref{thm:main}.
}

\updt{
\subsection{Proof of Theorem~\ref{thm:event}}
 The proof of Proposition~\ref{prop:event} remains valid due to the fact that~$d>0$ is sub-linear in $n$, and we let~$n \to \infty$. Specifically, by the pigeonhole principle it is clear that there must exist a sub-sequence of $\s{Z}^{\pi}$, of length at least $n/d$, which consists of consecutive $\s{X}_i$'s; therefore, we can apply the proof of Proposition~\ref{prop:event} on this sub-sequence. In this case, we note that the range of the parameter $p$ in the proof of Proposition~\ref{prop:event} becomes $(p+1)k\leq \left\lfloor  n/d \right\rfloor$, and thus as long as $n/d\to\infty$ we have $p\to\infty$ as well. Thus, the conclusion of Proposition~\ref{prop:upper} remains valid. The bound in~\eqref{eq:uperrbnd} also holds, due to the condition in~\eqref{eq:ybound}. Thus,~\eqref{eq:ulbnd} holds, and this completes the proof of Theorem~\ref{thm:event}.~\hfill{$\square$}
}


\updt{
\subsection{Proof  of Theorem~\ref{thm:event2}}
 As in the proof of Theorem~\ref{thm:main}, define $\s{W}_i : = \s{X}_i^{-1/2}$,  $i\in\{0,1,\ldots,n\}$. The proof of the upper bound in Theorem~\ref{thm:event2} is in fact the same as in \eqref{eq:uperrbnd}. Indeed, in \eqref{eq:uperrbnd} we show that
\begin{align}
\lambda_{\s{max}}(\s{T}_n) &\leq\max_{i \in \{1,\dots,n\} } ( \s{W}_{i-1}+\s{W}_{i}), 
\end{align}
which implies the lower bound in \eqref{eqn:Rgeneral}. The upper bound in \eqref{eqn:Rgeneral} follows from the same arguments used to obtain Proposition~\ref{prop:event}. Indeed, for any $1\leq k\leq n$, let $\s{I}_k$ be any set of $k$ consecutive indices in $\{0,1,\ldots,n\}$. Let $\s{P}_n$ be the matrix obtained by replacing the~$(k+1)\times (k+1)$ principal minor that corresponds to the indices $\s{I}_k$ 
of $ \s{T}_n$ by~$\s{M}_k\cdot \min_{i\in\s{I}_k}\s{W}_i$. Note that~$ \s{T}_n \geq \s{P}_n$ (where the inequality is componentwise), and thus~$\lambda_{\max} (\s{T}_n ) \geq \lambda_{\max} (\s{P}_n )$. By  Cauchy's interlacing theorem, the largest eigenvalue of~$\s{P}_n$
is larger or equal to the largest eigenvalue of any of its principal minors. Thus,
\begin{align}
\lambda_{\max} (\s{P}_n	)&\geq \lambda_{\max} \left(\s{M}_k\cdot \min_{i\in\s{I}_k}\s{W}_i\right) \nonumber \\
& \geq 2\min_{i\in\s{I}_k}\s{W}_i \cdot\cos\left({\frac{\pi}{k+2}}\right).\label{eqn:lowerBound:general}
\end{align}
Now, since \eqref{eqn:lowerBound:general} holds for any choice of $1\leq k\leq n$ and $\s{I}_k\in\mathcal{I}_k^{n+1}$, we can maximize the r.h.s. of \eqref{eqn:lowerBound:general} with respect to these assignments, which implies the upper bound in \eqref{eqn:Rgeneral}.
~\hfill{$\square$}
}

\subsection*{Acknowledgements}
The authors thank Yoram Zarai for  helpful comments. The work of MM is partially supported by a research grant from  the Israel Science Foundation~(ISF).
\updt{We thank the anonymous
reviewers and the editor for many helpful comments and for the timely review process.}


 \bibliographystyle{abbrv}
 \bibliography{RFM_bibl_wasim}
 
\end{document}